\documentclass[twocolumn, trackchanges]{aastex631}

\usepackage{amsmath}
\newcommand{\dif}{\mathrm{d}}

\usepackage[normalem]{ulem}

\begin{document}

\title{On the potential cosmogenic origin of the ultra-high-energy event KM3-230213A}
\email{km3net-pc@km3net.de, acondorelli@km3net.de, antonio.marinelli@na.infn.it}
\collaboration{350}{The KM3NeT Collaboration}
\author{O.~Adriani}
\affiliation{INFN, Sezione di Firenze, via Sansone 1, Sesto Fiorentino, 50019 Italy}
\affiliation{Universit{\`a} di Firenze, Dipartimento di Fisica e Astronomia, via Sansone 1, Sesto Fiorentino, 50019 Italy}
\author{S.~Aiello}
\affiliation{INFN, Sezione di Catania, (INFN-CT) Via Santa Sofia 64, Catania, 95123 Italy}
\author{A.~Albert}
\affiliation{Universit{\'e}~de~Strasbourg,~CNRS,~IPHC~UMR~7178,~F-67000~Strasbourg,~France}
\affiliation{Universit{\'e} de Haute Alsace, rue des Fr{\`e}res Lumi{\`e}re, 68093 Mulhouse Cedex, France}
\author{A.\,R.~Alhebsi}
\affiliation{Khalifa University of Science and Technology, Department of Physics, PO Box 127788, Abu Dhabi,   United Arab Emirates}
\author{M.~Alshamsi}
\affiliation{Aix~Marseille~Univ,~CNRS/IN2P3,~CPPM,~Marseille,~France}
\author{S.~Alves~Garre}
\affiliation{IFIC - Instituto de F{\'\i}sica Corpuscular (CSIC - Universitat de Val{\`e}ncia), c/Catedr{\'a}tico Jos{\'e} Beltr{\'a}n, 2, 46980 Paterna, Valencia, Spain}
\author{A.~Ambrosone}
\affiliation{Universit{\`a} di Napoli ``Federico II'', Dip. Scienze Fisiche ``E. Pancini'', Complesso Universitario di Monte S. Angelo, Via Cintia ed. G, Napoli, 80126 Italy}
\affiliation{INFN, Sezione di Napoli, Complesso Universitario di Monte S. Angelo, Via Cintia ed. G, Napoli, 80126 Italy}
\author{F.~Ameli}
\affiliation{INFN, Sezione di Roma, Piazzale Aldo Moro 2, Roma, 00185 Italy}
\author{M.~Andre}
\affiliation{Universitat Polit{\`e}cnica de Catalunya, Laboratori d'Aplicacions Bioac{\'u}stiques, Centre Tecnol{\`o}gic de Vilanova i la Geltr{\'u}, Avda. Rambla Exposici{\'o}, s/n, Vilanova i la Geltr{\'u}, 08800 Spain}
\author{L.~Aphecetche}
\affiliation{Subatech, IMT Atlantique, IN2P3-CNRS, Nantes Universit{\'e}, 4 rue Alfred Kastler - La Chantrerie, Nantes, BP 20722 44307 France}
% ...
\begin{abstract}

On the 13th February 2023 the KM3NeT/ARCA telescope observed a track-like event compatible with a ultra-high-energy muon with an estimated energy of 120 PeV, produced by a neutrino with an even higher energy, making it the most energetic neutrino event ever detected. A diffuse cosmogenic component is expected to originate from the interactions of ultra-high-energy cosmic rays with ambient photon and matter fields. The flux level required by the KM3NeT/ARCA event is however in tension with the standard cosmogenic neutrino predictions based on the observations collected by the Pierre Auger Observatory and Telescope Array over the last decade of the ultra-high-energy cosmic rays above the ankle (hence from the local Universe, $z\lesssim 1$). We show here that both observations can be reconciled by extending  the integration of the equivalent cosmogenic neutrino flux up to a redshift of $z\simeq 6$ and assuming a subdominant fraction of protons in the ultra-high-energy cosmic-ray flux, thus placing constraints on known cosmic accelerators. 

\end{abstract}



\section{Introduction} 
\label{sec:intro}
\begin{figure}[h!]
    \centering
    \includegraphics[width=0.4\textwidth]{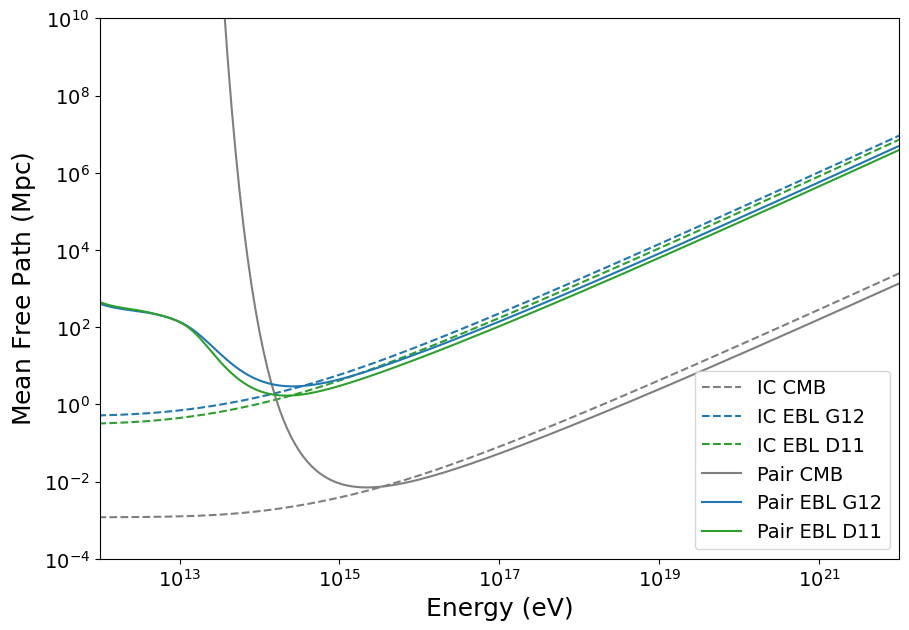} 
        \includegraphics[width=0.42\textwidth]{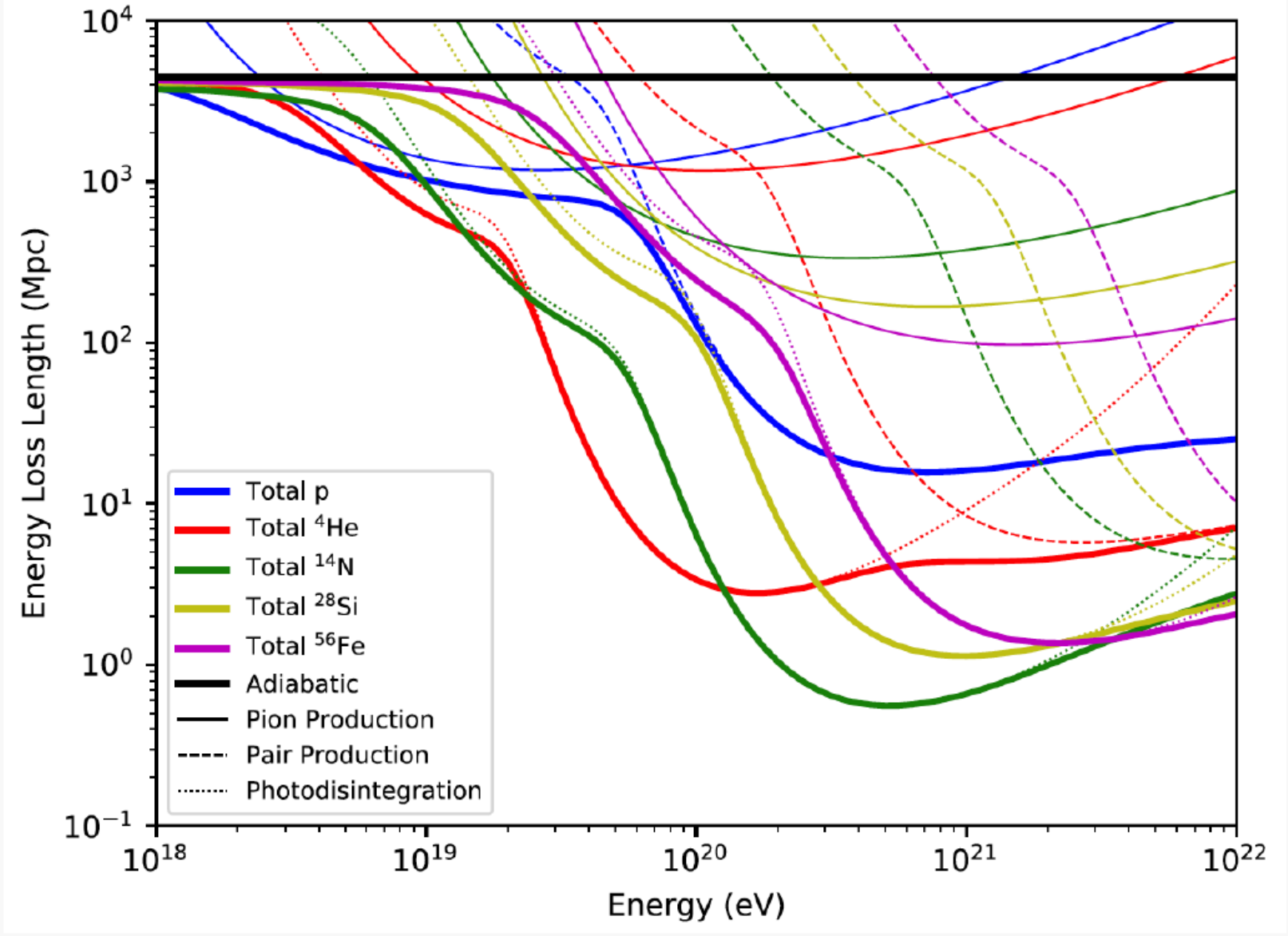} 

    \caption{Top: mean free path of photons as a function of energy. Pair production (Pair) and inverse-Compton (IC) interactions with photons of the CMB and EBL (for two different models) are shown.
    Bottom: Energy loss length  of proton and nuclei as a function of energy. Different interactions with photons of the CMB and EBL are considered. The plots have been made using the CRPropa \citep{AlvesBatista:2022vem} software.}
    \label{fig:photon_prop}
\end{figure}

The presence of extreme cosmic accelerators in our Universe has been 
established in the last decades by the observation of gamma rays, cosmic rays and 
neutrinos~\citep{IceCube:2013low,Fermi-LAT:2009ihh,PierreAuger:2015eyc}. A multi-messenger approach aims to combine detections of various cosmic 
messengers to overcome observational limitations and unveil the underlying 
astrophysical processes in a source environment or in the extragalactic medium.\\ However, the distant Universe remains unexplored at the highest energies because it is opaque to both cosmic rays and gamma rays.
Gamma rays with energies exceeding a few TeV interact with low-energy photons of the extragalactic background light (EBL)~\citep{Dominguez:2010bv,Finke:2009xi} or cosmic microwave background (CMB)~\citep{ Planck:2018vyg}, resulting in $e^+\,e^-$ pair production, which severely attenuates their propagation over cosmological distances, as shown in the top panel of figure \ref{fig:photon_prop}.
Similarly, cosmic rays at the highest energies ($\geq 10^{18} \ \rm eV$) suffer energy losses and photo-disintegration through interactions with CMB and EBL photons~\citep{Berezinsky:2002nc}, resulting in a mean free path smaller than a few Gpc (which corresponds to a redshift $z\simeq 1$), as shown in the bottom panel of figure \ref{fig:photon_prop}. At lower energies, charged nuclei are also deflected by the ambient magnetic fields, which further impedes the identification of their sources.\\
In contrast to the other messengers, neutrinos interact only weakly with matter and radiation, allowing them to travel virtually unimpeded over cosmological distances. This makes neutrinos invaluable probes of the most remote and densest regions of the Universe, where other messengers are absorbed or scattered. The detection of ultra-high energy (UHE) neutrinos could therefore constitute a unique proof of the presence of extreme accelerators in the far-away Universe.\\
This work investigates the possibility that the neutrino event (KM3-230213A) observed in KM3NeT/ARCA with an energy in the 90\%  confidence level (CL) range (72 PeV $-$ 2.6 EeV) is a cosmogenic neutrino produced in a region of the Universe deep enough to account for the estimated neutrino flux.\\
{In \citep{KM3NeT2024}, a wide range of cosmogenic models has been presented (see Table 4 of \citep{KM3NeT2024}): some  minimal scenarios \citep{Boncioli:2018lrv, Berat:2023ttm, Condorelli:2022vfa,Heinze:2019jou} predict a very low cosmogenic neutrino flux.
Other works, instead,  result in more optimistic expectations \citep{Aloisio:2015ega,PierreAuger:2022atd, Ehlert:2023btz,Muzio:2023skc, Zhang:2018agl} by considering, for example, the interplay between two source populations.
In this work, a fit of cosmic-ray data at the highest energies is performed, illustrating how the expected flux can be enhanced starting from a minimal scenario. }\\
The paper is organized as follows: In section~\ref{sec:cosmo_neutrinos} cosmogenic neutrinos and their importance for multi-messenger astronomy are described.  In section~\ref{sec:Evt},  KM3-230213A as observed by the KM3NeT/ARCA detector is detailed. Using the inputs of sections ~\ref{sec:cosmo_neutrinos} and ~\ref{sec:Evt}, the adopted cosmological scenario is reported in section~\ref{sec:framework} and the spectra produced in the nearby and far-away Universe are compared in section \ref{sec:results}. Finally, the significance of the results is discussed in section~\ref{sec:discussion}. 

\section{Cosmogenic neutrinos}
\label{sec:cosmo_neutrinos}
The energy spectrum of cosmic rays has been observed to extend beyond $10^{20}~\mathrm{eV}$~\citep{Linsley:1963km}, indicating that charged particles can be accelerated to ultra-high energies in powerful astrophysical objects, although their exact sources remain unidentified. Through interactions in their source environments or during their journey to Earth, cosmic rays produce neutrinos with energies corresponding to a fraction of their energy, hereinafter called cosmogenic neutrinos. 
UHE neutrinos are predicted to arise from the interactions of cosmic rays with background photon fields permeating the Universe. This mechanism, extensively studied in the literature~\citep[e.g.,][]{Hill:1983mk,Protheroe:1995ft,Lee:1996fp,Waxman:1998yy,Engel:2001hd,Ahlers:2010fw,Kampert:2011hkm,PierreAuger:2022atd,PierreAuger:2023mid}, relies on several assumptions, leading to flux predictions that can vary by orders of magnitude. The dominant production channel involves the decay of $\pi^\pm$ mesons, created by primary proton cosmic rays or secondary nucleons produced during the photo-disintegration of nuclei interacting with background photons. As nucleons inherit a fraction of the fragmented nucleus energy, neutrinos from heavier nuclei tend to have lower energies than those from lighter nuclei or protons. Consequently, the cosmogenic neutrino flux is strongly dependent on the cosmic-ray mass composition, which is poorly constrained above $5 \times 10^{19}~\mathrm{eV}$ \citep{PierreAuger:2023bfx}. Other key factors include the maximum acceleration energy, the shape of the particle energy spectrum, and the cosmological evolution of their sources. Advances in constraining these parameters have significantly altered flux predictions at $10^{18}~\mathrm{eV}$: initial estimates of $E^2\phi_\nu \sim 10^{-9}~\mathrm{GeV~cm^{-2}~sr^{-1}~s^{-1}}$, assuming a pure-proton composition, have dropped to $E^2\phi_\nu \sim 10^{-12}~\mathrm{GeV~cm^{-2}~sr^{-1}~s^{-1}}$ under mixed-composition models, consistent with data from the Pierre Auger Observatory~\citep{PierreAuger:2014gko,PierreAuger:2014sui,PierreAuger:2023xfc} and other experiments~\citep{Watson:2021rfb}.


Finally, cosmogenic neutrinos can also be produced through interactions between UHE particles and interstellar matter in the Galactic disk. This mechanism mirrors the production of lower-energy neutrinos recently observed from Galactic cosmic rays~\citep{IceCube:2023ame}. Interactions of Ultra-High-Energy cosmic rays (UHECR) within our Galaxy provide a guaranteed baseline for cosmogenic neutrino production \citep{2024ApJ...966..186B}.\\
The procedure to compute cosmogenic neutrino fluxes is detailed in the  section \ref{sec:framework}.

\section{KM3$-$230213A}
\label{sec:Evt}
An UHE muon traversing the KM3NeT/ARCA detector was observed on February 13,
2023 at 01:16:47 UTC (KM3-230213A). At that time, 21 detection units were in operation. The detector, referred to as ARCA21, collected data in this configuration from September 23, 2022 until September 11, 2023, for a total livetime $T^{\rm ARCA21} $ of 287.4 days. Over this period, about 110 million events were triggered,
with KM3$-$230213A being the highest-energy event observed. The estimated muon energy is $120^{+110}_{-60}$ PeV, with a $90\%$ C.L. interval of 35–380 PeV, while the corresponding median neutrino energy to produce such a muon in the simulations of the ARCA detector is 220 PeV; the 68\% (90\%) of simulated events fall in the 110$–$790 PeV (72 PeV$–$2.6 EeV) energy range. 

 \section{Description of Cosmogenic scenario}
 \label{sec:framework}

 \subsection{Injected cosmic rays composition}
The non-thermal processes responsible for accelerating the different types of particles are usually modeled using power-law spectra, while an exponential suppression is used to describe the end of the acceleration process in the absence of any strong indication from theory. Sources are believed to accelerate diverse proportions of nuclei, which, for simplicity, are  grouped into five stable representative nuclei: hydrogen ($^1$H), helium ($^4$He), nitrogen ($^{14}$N), silicon ($^{28}$Si), and iron ($^{56}$Fe).
The ejection rate $q_{{A}}(E)$ of nuclei with mass number ${A}$ per comoving unit volume and per unit energy of nucleons is usually modeled as 
\begin{equation}
\label{eqn:qA}
    q_{{A}}(E) = q_{0{A}}\left(\frac{E}{E_0}\right)^{-\gamma_{A}}f(E,Z_{{A}}),
\end{equation}
where $\gamma_{A}$ is the spectral index and $q_{0{A}}$ are the injection rates. The suppression function used for nucleons and nuclei is the same as in the reference case of \citep{PierreAuger:2016use}:
\begin{equation}
\label{eqn:fsupp}
    f(E,Z) = 
    \begin{cases}
    1 & \mathrm{if~}E\leq E^{Z}_{\mathrm{max}},\\
    \exp{\left(1-E/E^{Z}_{\mathrm{max}}\right)} & \mathrm{otherwise}.
    \end{cases}
\end{equation}
The maximum acceleration energy is expected to be proportional to the electric charge of each element, $E^{Z}_{\mathrm{max}} = ZE_{\mathrm{max}}$, where $E_{\mathrm{max}}$ is  a single free parameter.

For each atomic species $A$, the differential energy production rate per comoving volume unit of the sources, which is directly connected to their differential luminosity, is consequently $\ell_A(E,z)= E^2q_A(E)S(z)$, where $S(z)$ reflects the redshift evolution of the UHECR luminosity density. The bolometric energy production rate per comoving volume unit at redshift $z$, on the other hand, is calculated as
\begin{equation}
    \mathcal{L}_A(E, z)=S(z)\int_{E}^{\infty}\dif E'E'q_A(E').
\end{equation}
    
Its average value in a volume spanning in the range [$z_\text{min}, z_\text{max}$] is 
\begin{equation}
 \bar{\mathcal{L}}_A(E) = \int_{z_\text{min}}^{z_\text{max}} \dif z \left|\frac{\dif t}{\dif z}\right| \mathcal{L}_A(E, z) / \int_{z_\text{min}}^{z_\text{max}} \dif z \left|\frac{\dif t}{\dif z}\right|,   
\end{equation}
where $t(z)$ is the look-back time. In this scenario, since considering only injected particles above the so-called `ankle' ($10^{18.7} \  \rm eV$) are considered, the effects of the extragalactic magnetic field on particle propagation are neglected, allowing particle propagation to be effectively treated as one-dimensional \citep{Mollerach:2013dza}.
The energy spectrum is described using $\log_{10}(E/\rm eV)$ bins with a width of 0.1, spanning from ${17.8}$ to ${20.2}$. This measurement is based on 15 years of data collected with the surface detector array of the Pierre Auger Observatory \citep{PhysRevLett.125.121106}. To present the $X_{\rm max}$ distributions \citep{PRDpaper2014}, $\log_{10}(E/\rm eV)$ bins of 0.1 from ${17.8}$ to $19.6$  are used, with an additional, broader bin including events with energies above $10^{19.6} \ \rm eV$. Each $X_{\rm max}$ distribution is divided into intervals of 20 $ \rm g \ cm^{-2}$.
The agreement between the model described in this paper and the UHECR data is evaluated following the procedure outlined in \citep{combFit}.

\subsection{Cosmological sources evolution}
The bulk of UHE cosmic rays detected by the Pierre Auger Observatory \citep{PierreAuger:2021hun} and by the Telescope Array \citep{ABBASI2023102864}  as well as their measured spectrum are mostly driven by closeby astrophysical accelerators, implying that UHE-related source properties are mostly unexplored for $z\geq 1$. Different astrophysical candidates, steady or transients, can accelerate cosmic rays up to $10^{21}$ eV when extreme environments are considered; among them different classes of Active Galactic Nuclei (AGN)~\citep{Blandford:2018iot,Berezinsky:2002nc}, Starburst Galaxies (SBGs)~\citep{Bykov:2020zqf}, Gamma-ray Bursts (GRB)~\citep{Vietri:1995hs}, and Tidal Disruption Events~\citep{Guepin:2017abw,Plotko:2024gop}.
Their comoving source density is generally parametrized as:
\begin{equation}S(z) \propto (1+z)^{m}
\label{eqn:cosmditr}
\end{equation}
In this work, $m$ is varied from $-5$ to $5$ with a step of 0.2, covering all possible source evolution scenarios without assuming a specific source type. At low redshifts ($z \lesssim 1$), a strong positive evolution ($m = 5$) could correspond to high-luminosity sources, such as jetted objects, e.g., blazars, observed in gamma rays~\citep{Ajello2014}, or non-jetted sources like high-luminosity Seyfert galaxies~\citep{Ueda2014}. A weaker positive evolution ($m = 3$) aligns with the star formation rate (SFR) evolution~\citep{Madau2014}. Flat evolution ($m = 0$) may be linked to the stellar-mass density in the Universe~\citep{Fukazawa2022}, low-luminosity Seyfert galaxies in X-rays~\citep{Kochanek2016}, or intermediate-luminosity BL Lacs and FSRQ~\citep{Ajello2014}. Negative evolutions ($m = -3$) are associated with low-luminosity BL Lacs~\citep{Ajello2014}, radio galaxies~\citep{Fukazawa2022}, or the redshift evolution of tidal disruption events~\citep{Kochanek2016}. At higher redshifts ($z \gtrsim 1$), the evolution of some of these source classes remains uncertain.\\
The resulting flux of cosmogenic neutrinos can vary with distance, depending on the assumed source evolution. The emission rate density $L(E,z)$ of cosmic rays per comoving volume for different accelerator populations can be expressed as:

\begin{equation}
L(E,z) = S(z) \times Q_{CR}(E),
\label{eqn:Lumin}
\end{equation}
where $Q_{CR}(E)$ represents the injection term for the source type. The implications of this assumption and its associated uncertainties are discussed in Section~\ref{sec:discussion}.\\

 \section{Expected neutrino fluxes for different redshift}
  \label{sec:results}
  \begin{figure}[t]
    \centering
    \includegraphics[width=1.25\linewidth]{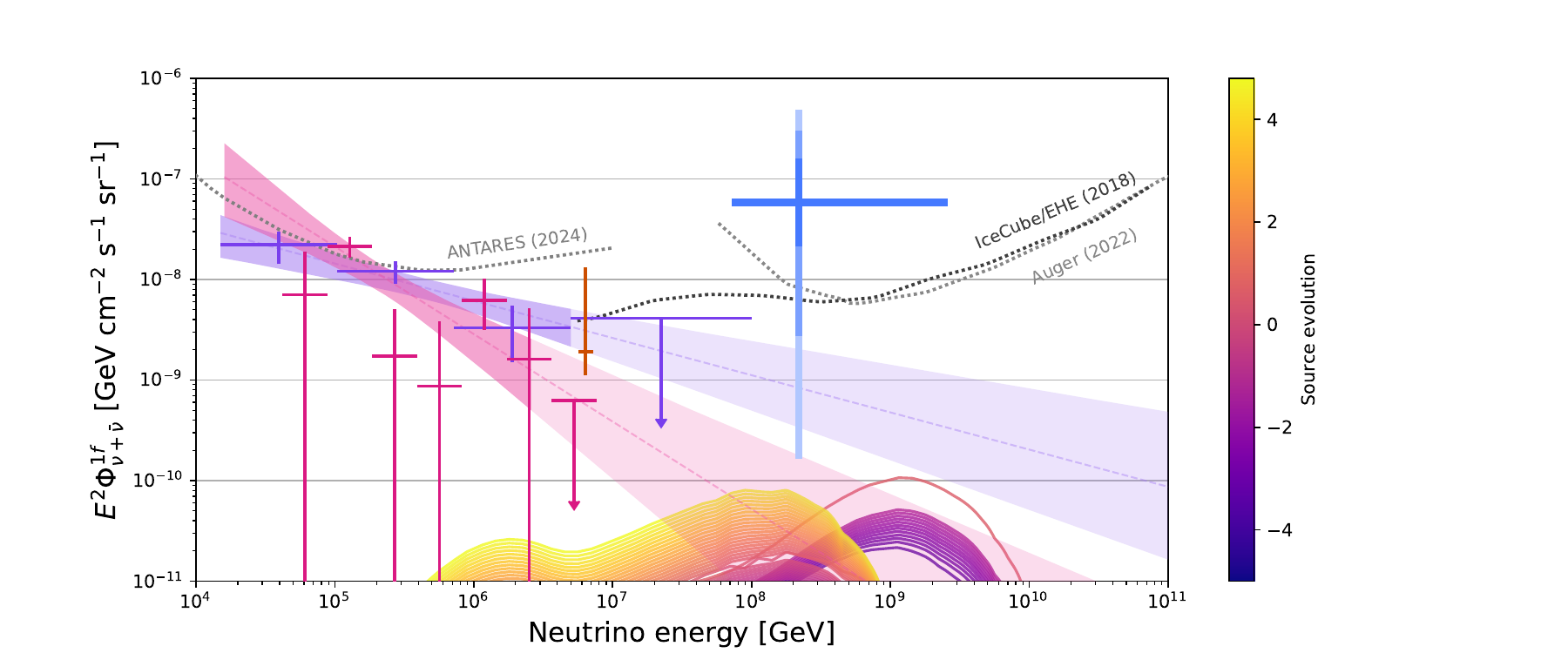} 
        \includegraphics[width= 1.25 \linewidth]{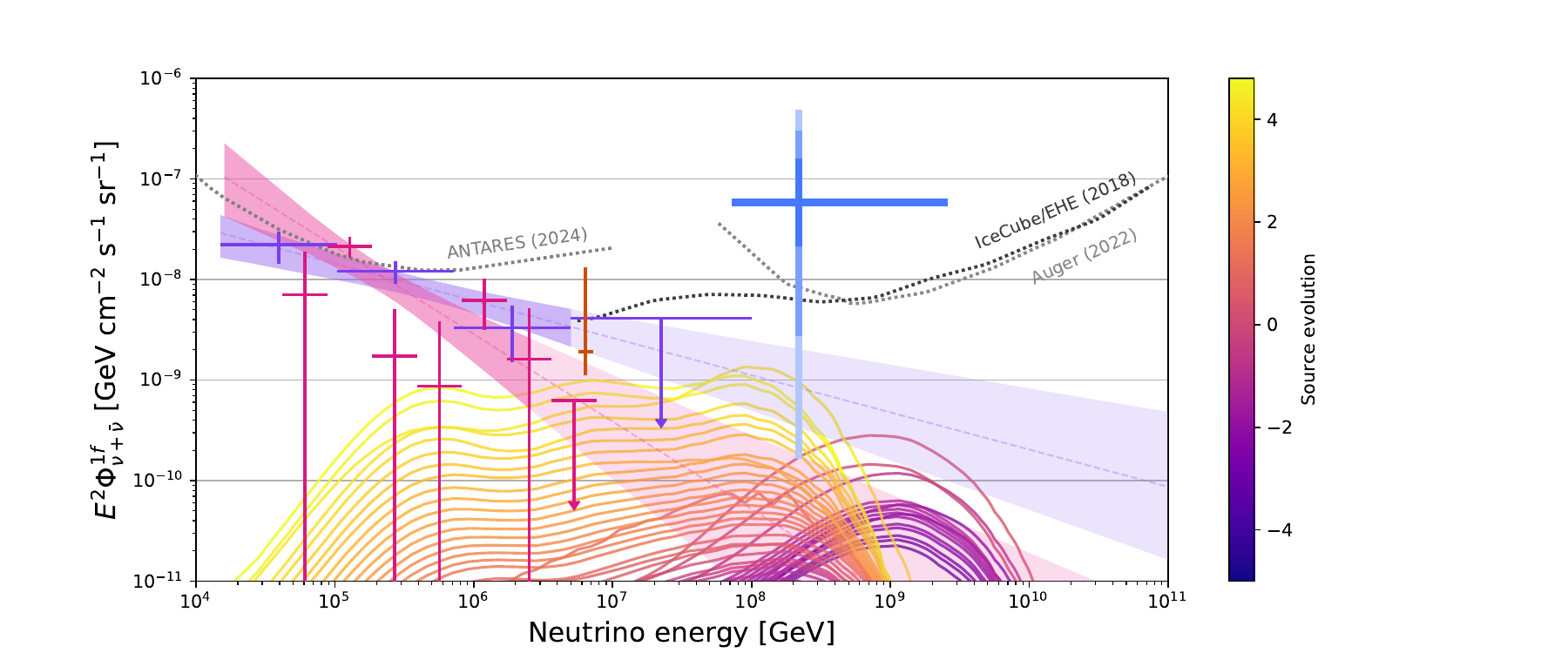} 

    \caption{Expected neutrino fluxes as a function of energy for different  source evolutions (colour code) for two different maximum redshift values: $z_{\rm max} = 1$ (top panel) and $z_{\rm max} = 6$ (bottom panel). \\
    The blue cross corresponds to the flux needed to produce one expected event in the central 90\% CL range of neutrino energy associated with the KM3-230213A event (horizontal span); the vertical bars represent the 1, 2 and 3$\sigma$ Feldman-Cousins confidence intervals on this estimate \citep{KM3NeT2024}. The purple- and pink-filled regions represent the 68\% confidence level contours of the IceCube single power-law fits (Northern-Sky Tracks, and High-Energy Starting Events, respectively):
the darker-shaded regions are the respective 90\% central energy range at the best fit (dashed line), while the lighter-shaded regions are extrapolations to higher energies. The purple and pink crosses are the fit from the same analyses, while the orange cross corresponds to the
IceCube Glashow resonance event. The dotted lines are upper limits from ANTARES (95\%
confidence level), Pierre Auger and IceCube.}
    \label{fig:candidate}
\end{figure}

Using this framework, a scan over plausible source evolutions, parameterized as in equation \ref{eqn:cosmditr}
with \( m \) ranging from \(-5\) to \(5\) has been performed. The cosmogenic neutrino fluxes associated to the best fit parameters that describe UHECR energy spectrum and composition are shown in figure~\ref{fig:candidate}. In the top panel, we limit the UHECR source distribution up to redshift \( z_{\rm max} = 1 \), which represents the distance beyond which UHECRs above the so-called ankle are expected to interact significantly. In the bottom panel, the source distribution has been extended up  to redshift \( z_{\rm max} = 6 \), considering that the Universe is opaque to high-energy neutrinos at early times~\citep{Berezinsky:1991aa,Gondolo:1991rn}. \\
In both cases, the neutrino fluxes exhibit two distinct bumps in the energy range of interest, corresponding to interactions with the EBL at lower energies and the CMB at higher energies. {A noticeable change of shape can be observed when transitioning from negative to positive source evolution, as reflected in the plots, where violet tones predominantly represent the former and yellow tones the latter.} 
This effect arises from transitioning between regions of parameter space where a negligible quantity of protons is predicted to configurations that are proton-rich. Negative source evolution implies that the total flux is dominated by nearby sources, which provides less room for photodisintegration of heavy nuclei. Consequently, the fit favours an intermediate mass composition at the source, closely resembling what is observed at Earth. In contrast, a positive evolution parameter suggests that most sources are farther away, allowing more photodisintegration processes. This enables a heavier composition at the source, which becomes lighter at Earth due to interactions in the interstellar medium. 
{As a result, the neutrino flux associated with positive source evolution is more abundant in both cases of $z_{\rm max} = 1$ and $z_{\rm max} = 6$.}
Along this process, a significant number of protons are produced, which play a critical role in neutrino production, as will be further discussed in the following section.\\

The two panels of figure \ref{fig:candidate} show that for the \( z_{\rm max} = 1 \) scenario, the fluxes remain below \( 10^{-10} \, \rm GeV \, cm^{-2} \, s^{-1} \, sr^{-1} \) in the energy of interest. In contrast, considering sources at higher redshifts (\( z_{\rm max} = 6 \)) can lead to an increase of the cosmogenic flux by about one order of magnitude, suggesting that, if the observed neutrino is cosmogenic, it is more likely to originate from distant sources rather than being associated with the UHECR flux measured above the ankle.\\
The expected number of events $n_{\rm exp}$ for a specific flux model $\Phi_{\rm model}$ is evaluated as:
\begin{equation}
    \begin{split}
        n_{\rm exp}^{\rm ARCA21} &= 
        T^{\rm ARCA21} \sum_i 
        \int_{\Delta\Omega_i} \int_{\Delta E} 
        A_{\rm eff}^{\rm ARCA21}(E, \Omega) \,  \\
        &\quad \times \Phi_{\rm model}(E, \Omega) \, 
        {\rm d}E \, {\rm d}\Omega.
    \end{split}
    \label{eq:eff_modified}
\end{equation}
where $A_{\rm eff}^{\rm ARCA21}$ represents the effective areas for events in different ranges of zenith angle $\Delta\Omega_i$.
For each flux model, the expected event rate in ARCA21 is calculated by integrating over the full energy range as in \citep{KM3NeT2024}, including the region above \( 100 \, \mathrm{PeV} \). The effective area is computed  using the same selections as adopted in \citep{KM3NeT2024}.

\begin{figure}[t]
    \centering
    \includegraphics[width=\linewidth]{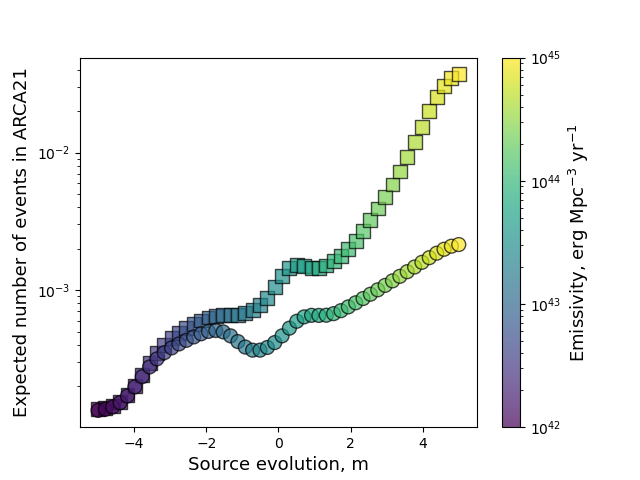} 

    \caption{Expected number of events in ARCA21 as a function of source evolution for $z_{\rm max} =1$ (circles) and $z_{\rm max} = 6$ (squares). The colour axis represents the emissivity of plausible cosmic-ray sources.}
    \label{fig:evol}
\end{figure}

The computed neutrino fluxes can be then translated into expected number of events. Using equation~\ref{eq:eff_modified}, the expected number of events has been evaluated for various source evolutions, considering two scenarios: \( z_{\rm max} = 1 \) and \( z_{\rm max} = 6 \). The results are shown in figure~\ref{fig:evol}: for negative values of \( m \), the two scenarios are nearly indistinguishable due to the very low neutrino flux. However, as \( m \) increases, the distinction between nearby and distant Universe scenarios becomes evident, with significantly more events expected in the \( z_{\rm max} = 6 \) case. \\
Furthermore, the parameter space regions that best describe our observations are those with higher \( m \) values, indicating higher emissivity (see colour axis in figure~\ref{fig:evol}). This result is expected, as positive \( m \) values imply more interactions and, consequently, the power of the accelerators must be higher in these scenarios.\\
Different extragalactic objects can be identified as UHECR accelerators considering their physical properties mainly derived through electromagnetic observations. Even though such properties can be assumed to remain similar for a significant fraction of the history of the Universe~\citep{Jakobsson:2005jc}, it is important to probe the efficiency and cosmological evolution of UHE accelerators up to high redshift through UHE astrophysical messengers. The observation of KM3-230213A gives  the possibility to explore a new region of the Universe in the energy range of $10^{17}-10^{18}$ eV for $\nu$ (therefore up to $10^{20} \ \rm eV$ for cosmic rays) so far untested by the Pierre Auger and Telescope Array observatories. 
It would be valuable to compare the emissivity values reported in the colour axis of figure \ref{fig:evol} with those of plausible source candidates. However, it is crucial to emphasize that the reported values represent emissivity in {cosmic rays}, while source emissivity is typically expressed in terms of gamma rays or electromagnetic emission. As a result, a direct comparison is not straightforward without applying a conversion factor to estimate the fraction of energy channeled into cosmic rays relative to the electromagnetic component.\\

\section{Constraining the proton fraction at the highest energies}
Protons are significantly more effective than nuclei in producing pions and high-energy neutrinos, as they interact efficiently with photon fields (primarily the CMB at the highest energies) through photo$-$pion production (\(p + \gamma \rightarrow \Delta^+ \rightarrow \pi + N\)), generating charged pions that decay into neutrinos.
 The photo-pion production process holds also for nucleons bound within UHE nuclei, being the interacting nucleon ejected from the parent nucleus, but this process is subdominant with respect to nucleus photo-disintegration except at extremely high energies. Thus, with respect to nuclei, protons dominate in pion production, and therefore in neutrino production, at high energies.\\
 Recent works \citep{PierreAuger:2022atd, Ehlert:2023btz} have demonstrated that the inclusion of a proton component, while preserving the expected mass composition (predominantly intermediate masses), can significantly influence the predicted neutrino flux.\\
This scenario has been explored within the cosmogenic framework by introducing a secondary component, made only by protons, with a different spectral index, in order to fit the proton fraction mass composition as measured by the Pierre Auger Observatory. This was implemented by fitting the energy spectrum and mass composition data below the ankle, using the proton spectrum measured by Auger, and extrapolating it to the highest energies.

\begin{figure}[h!]
    \centering
    \includegraphics[width=0.5\textwidth]{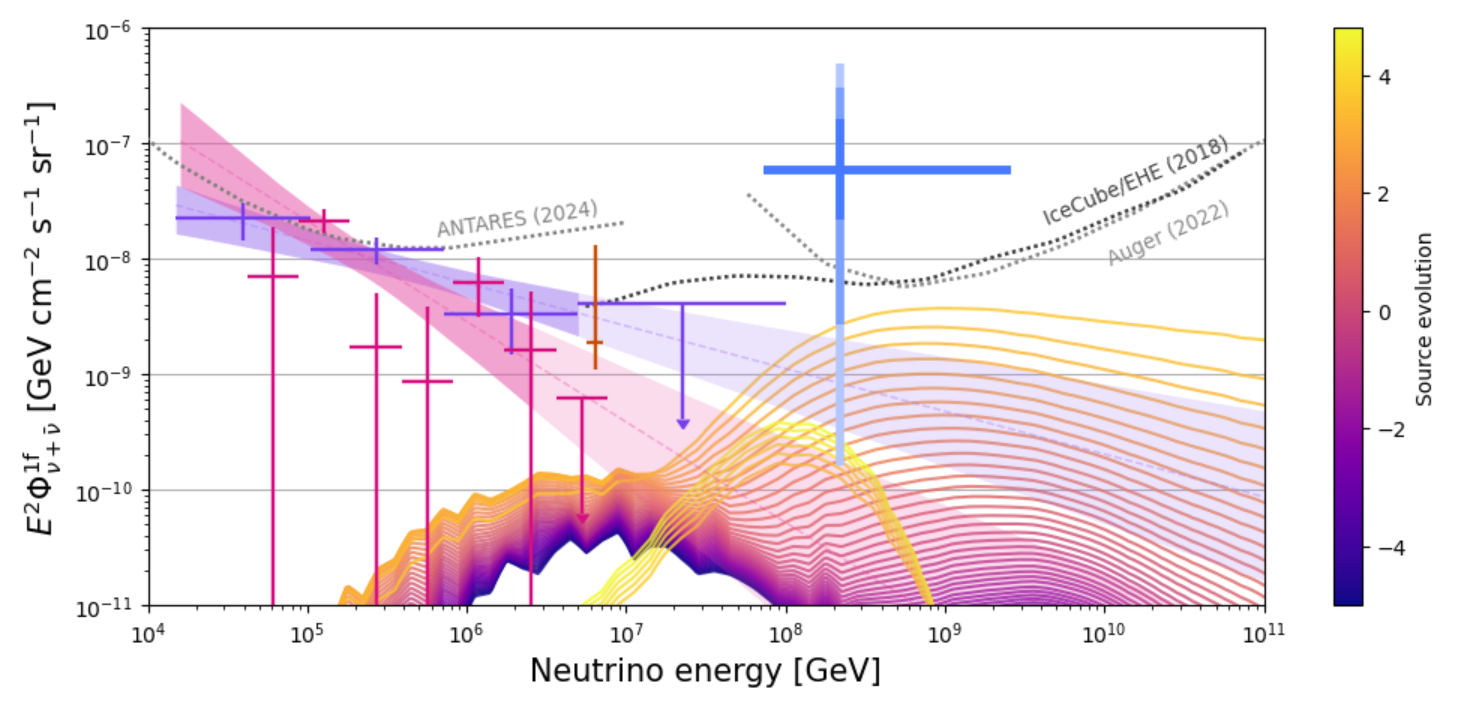} 
    \caption{Expected neutrino fluxes as a function of energy assuming a second component, made of protons, subdominant at the highest energies. The figure follows the same format as figure \ref{fig:candidate}.}
    \label{fig:proton}
\end{figure}

The resulting neutrino flux predictions are shown in figure~\ref{fig:proton}, highlighting how the presence of a subdominant proton component can enhance the expected neutrino flux. 

In conclusion, even a small fraction of protons at the highest energies translates into a significant increase of the cosmogenic neutrino flux, which can offer prospects for more future detections~\citep{Rodrigues:2020pli,PierreAuger:2022atd,Muzio:2023skc}.
The observation of KM3-230213A supports the possibility of a subdominant proton component at the highest energies, assuming the cosmogenic hypothesis.
\section{Discussion}
\label{sec:discussion}
The observation of KM3-230213A has triggered the investigation of its possible origin as a cosmogenic neutrino, based on the existence of known cosmic accelerators and on the available observations of UHECRs. The cosmogenic origin is compatible (0.06 expected events for the most optimistic scenario for the lifetime of ARCA21, which corresponds to a probability of 5.6 \% to observe an event) with the reported neutrino flux when looking into the deep sky ($z_{\rm max}\simeq 6$) and at strong positive cosmological source evolution, allowing for a non-negligible proton fraction at the highest energies.
\\
{In this work, results have been presented using a simple source evolution $S(z) \propto (1 + z)^m$ with constant $m$ across the entire redshift $z$ range. However, possible improvements to the model include exploring alternative assumptions, such as aligning the UHECR production rate in the Universe with the star formation rate \citep{Madau:2014bja}.} This idea stems from the premise that a higher rate of astrophysical explosions correlates with increased production of cosmic rays at the highest energies. \\
According to trends in literature, the SFR evolution is parametrized with a functional form, which features a rise at low redshifts, a plateau at intermediate redshifts, and a decline at high redshifts, following the parametric formula as reported in \citep{Aloisio:2015ega}. It is found that the neutrino flux associated to a population evolving as the SFR history is significantly lower with respect to the one obtained with a simplified source evolution shown in the bottom panel of figure \ref{fig:candidate}.
An analogous approach has been considered by assuming an AGN-like evolution \citep{Ahlers:2009rf}. Despite a more abundant flux ($\simeq 2$ times more abundant) with respect to the SFR one, the neutrino flux is still much lower than the ones shown in the previous section, using the evolution reported in equation \ref{eqn:cosmditr}. From these results it is possible to conclude that, if the origin of KM3-230213A is cosmogenic, the UHE production rate should not follow the distribution of matter so far known. Conversely, it also implies that a subdominant proton component is necessary to provide such abundant flux at the highest energies, irrespectively of the source evolution of the plausible sources. \\

Previous studies found that differences between EBL models had non-negligible impact in UHECR propagation \citep{AlvesBatista:2015jem}. Nonetheless, over the last years recent EBL models have been updated. Four of them have been tested, Gilmore \citep{Gilmore:2011ks}, Dominguez \citep{10.1111/j.1365-2966.2010.17631.x}, Saldana-Lopez \citep{2021MNRAS.507.5144S} and Andrews \citep{2018MNRAS.474..898A}. The uncertainties induced by the EBL models are found to be smaller than the systematic uncertainties on the energy spectrum and mass composition. Therefore, it can be concluded that the uncertainties on the EBL model do not influence the main message of this paper. \\

The impact of the photo-disintegration cross section model has also been investigated. Two different models have been used: PSB \citep{PSB} and \citep{Talys}, with the latter being employed in this analysis. It is observed that the differences in the proton fraction are on the order of 5\%, which has a negligible impact on the expected cosmogenic neutrino flux. Thus, these two sources of uncertainty in the modeling do not compromise the main conclusions of this work.\\

\section{Conclusion}
In this study, the cosmogenic origin of KM3-230213A has been investigated, with the aim of understanding their source features and the implications for high-energy astrophysical processes.
{In this work we show that, following the cosmogenic hypothesis, the theoretical models favor strong evolutions of the sources and a non-negligible proton fraction produced at the highest energies.} At the same time,  an additional diffuse extragalactic component is plausible, accounting for neutrinos produced in source environment. This aligns with the expectation that at the highest energies, the observed neutrino flux is not entirely cosmogenic; instead, it likely includes contributions from various astrophysical sources.\\
{In the future, KM3NeT sensitivity at the highest energies will enable more precise constraints on the expected cosmogenic fluxes, being competitive with the sensitivities achieved by other experiments \citep{IceCube:2022clp, PierreAuger:2019azx}}.\\
While uncertainties in UHECR source modeling persist, advancements in neutrino detection capabilities are expected to further further refine the understanding of UHE neutrino production. In doing so, they offer complementary insights that contribute to completing the multi-messenger astrophysics puzzle. Ultimately, this work highlights the potential of high-energy neutrino astronomy as a key probe for observing the most distant high-energy cosmic-ray accelerators, a task that will be further advanced by the capabilities of the KM3NeT telescope.
\section{Acknowledgments}

The authors acknowledge the financial support of:
KM3NeT-INFRADEV2 project, funded by the European Union Horizon Europe Research and Innovation Programme under grant agreement No 101079679;
Funds for Scientific Research (FRS-FNRS), Francqui foundation, BAEF foundation.
Czech Science Foundation (GAČR 24-12702S);
Agence Nationale de la Recherche (contract ANR-15-CE31-0020), Centre National de la Recherche Scientifique (CNRS), Commission Europ\'eenne (FEDER fund and Marie Curie Program), LabEx UnivEarthS (ANR-10-LABX-0023 and ANR-18-IDEX-0001), Paris \^Ile-de-France Region, Normandy Region (Alpha, Blue-waves and Neptune), France,
The Provence-Alpes-Côte d'Azur Delegation for Research and Innovation (DRARI), the Provence-Alpes-Côte d'Azur region, the Bouches-du-Rhône Departmental Council, the Metropolis of Aix-Marseille Provence and the City of Marseille through the CPER 2021-2027 NEUMED project,
The CNRS Institut National de Physique Nucléaire et de Physique des Particules (IN2P3);
Shota Rustaveli National Science Foundation of Georgia (SRNSFG, FR-22-13708), Georgia;

This work is part of the MuSES project which has received funding from the European Research Council (ERC) under the European Union’s Horizon 2020 Research and Innovation Programme (grant agreement No 101142396).
This work was supported by the European Research Council, ERC Starting grant \emph{MessMapp}, under contract no. $949555$.
The General Secretariat of Research and Innovation (GSRI), Greece;
Istituto Nazionale di Fisica Nucleare (INFN) and Ministero dell’Universit{\`a} e della Ricerca (MUR), through PRIN 2022 program (Grant PANTHEON 2022E2J4RK, Next Generation EU) and PON R\&I program (Avviso n. 424 del 28 febbraio 2018, Progetto PACK-PIR01 00021), Italy; IDMAR project Po-Fesr Sicilian Region az. 1.5.1; A. De Benedittis, W. Idrissi Ibnsalih, M. Bendahman, A. Nayerhoda, G. Papalashvili, I. C. Rea, A. Simonelli have been supported by the Italian Ministero dell'Universit{\`a} e della Ricerca (MUR), Progetto CIR01 00021 (Avviso n. 2595 del 24 dicembre 2019); KM3NeT4RR MUR Project National Recovery and Resilience Plan (NRRP), Mission 4 Component 2 Investment 3.1, Funded by the European Union – NextGenerationEU,CUP I57G21000040001, Concession Decree MUR No. n. Prot. 123 del 21/06/2022;
Ministry of Higher Education, Scientific Research and Innovation, Morocco, and the Arab Fund for Economic and Social Development, Kuwait;
Nederlandse organisatie voor Wetenschappelijk Onderzoek (NWO), the Netherlands;
The grant “AstroCeNT: Particle Astrophysics Science and Technology Centre”, carried out within the International Research Agendas programme of the Foundation for Polish Science financed by the European Union under the European Regional Development Fund; The program: “Excellence initiative-research university” for the AGH University in Krakow; The ARTIQ project: UMO-2021/01/2/ST6/00004 and ARTIQ/0004/2021;\\
Ministry of Research, Innovation and Digitalisation, Romania;
Slovak Research and Development Agency under Contract No. APVV-22-0413; Ministry of Education, Research, Development and Youth of the Slovak Republic;
MCIN for PID2021-124591NB-C41, -C42, -C43 and PDC2023-145913-I00 funded by MCIN/AEI/10.13039/501100011033 and by “ERDF A way of making Europe”, for ASFAE/2022/014 and ASFAE/2022 /023 with funding from the EU NextGenerationEU (PRTR-C17.I01) and Generalitat Valenciana, for Grant AST22\_6.2 with funding from Consejer\'{\i}a de Universidad, Investigaci\'on e Innovaci\'on and Gobierno de Espa\~na and European Union - NextGenerationEU, for CSIC-INFRA23013 and for CNS2023-144099, Generalitat Valenciana for CIDEGENT/2018/034, /2019/043, /2020/049, /2021/23, for CIDEIG/2023/20, for CIPROM/2023/51 and for GRISOLIAP/2021/192 and EU for MSC/101025085, Spain;\\
Khalifa University internal grants (ESIG-2023-008, RIG-2023-070 and RIG-2024-047), United Arab Emirates;\\
The European Union's Horizon 2020 Research and Innovation Programme (ChETEC-INFRA - Project no. 101008324).

\bibliography{sample631}{}
\bibliographystyle{aasjournal}



\end{document}